\documentclass[english,twocolumn,aps,pra,superscriptaddress,address,showpacs,showacknowledgments]{revtex4-2}
\usepackage[T1]{fontenc}
\usepackage[latin9]{inputenc}
\usepackage{color}
\usepackage{babel}
\usepackage{mathtools}
\usepackage{amsmath}
\usepackage{amssymb}
\usepackage{stackrel}
\usepackage{graphicx}
\usepackage[unicode=true,pdfusetitle,
 bookmarks=true,bookmarksnumbered=false,bookmarksopen=false,
 breaklinks=false,pdfborder={0 0 1},backref=false,colorlinks=false]
 {hyperref}
\hypersetup{
 colorlinks,linkcolor=blue,anchorcolor=blue,citecolor=blue,bookmarksnumbered}

\makeatletter
\usepackage{bm}
\usepackage{float}
\usepackage{amsthm}
\usepackage{mathrsfs}
\usepackage{xcolor}
\usepackage{babel}
\usepackage{hyperref}
\usepackage{textcomp}

\makeatother

\begin{document}
\title{Single-photon scattering by a giant molecule asymmetrically coupled
to parallel waveguides}
\author{Ze-Quan Zhang}
\affiliation{Fujian Key Laboratory of Quantum Information and Quantum Optics and
Department of Physics, Fuzhou University, Fuzhou 350116, China}
\author{Guang-Zheng Ye}
\affiliation{Fujian Key Laboratory of Quantum Information and Quantum Optics and
Department of Physics, Fuzhou University, Fuzhou 350116, China}
\author{Wei-Xin Chen}
\affiliation{Fujian Key Laboratory of Quantum Information and Quantum Optics and
Department of Physics, Fuzhou University, Fuzhou 350116, China}
\author{Yong Li}
\affiliation{Center for Theoretical Physics \& School of Physics and Optoelectronic
Engineering, Hainan University, Haikou 570228, China}
\author{Huaizhi Wu}
\affiliation{Fujian Key Laboratory of Quantum Information and Quantum Optics and
Department of Physics, Fuzhou University, Fuzhou 350116, China}
\begin{abstract}
We investigate single-photon scattering in a waveguide-QED
setup, where a giant molecule composed of two frequency-detuned
giant atoms are coupled to two parallel waveguides via multiple
connection points. The competition between coherent atom--atom coupling and the effective decay rates dictates the splitting of a single resonance into a doublet in the transmission (reflection) spectra. By tailoring the asymmetry of the decay rates and the atomic detuning, one can engineer photon-path interference to optimize the transfer between waveguides; under chiral coupling conditions, this interference can be further harnessed to realize fully deterministic routing. In the non-Markovian regime, retardation effects can reshape the spectra and actively drive transitions between the weak- and strong-coupling
regimes, converting an unsplit Markovian resonance into a clearly separated doublet, or conversely merging a split doublet back into a single resonance. For sufficiently long time delays, it further generates multiple resonances and avoided crossings, enriching the spectral response. Our results demonstrate how atomic detuning, decay-rate asymmetry, and non-Markovian retardation cooperate to
provide versatile, interference-based control over single-photon
routing in multi-port quantum networks.
\end{abstract}
\maketitle

\section{INTRODUCTION}

Waveguide quantum electrodynamics (QED) \citep{Gu2017,Roy2017,Sheremet2023}
studies the interaction between atoms (or other quantum emitters)
and propagating photons confined in waveguides, providing a versatile platform for controlling photon transport and implementing basic quantum-network
functionalities such as routing \citep{hoi2011demonstration,Hoi2012,Hoi2013,sundaresan2015beyond},
state transfer \citep{Wehner2018,Duan2010,Kimble2008}, and storage
\citep{Kimble2008,Roy2011}. In conventional waveguide QED, the size
of an atom is typically much smaller than the wavelength, so one can
adopt the dipole approximation and treat it as a point-like source
coupled to the waveguide at a single position \citep{Reiserer2015,Uppu2021,neumeier2013single,shen2007strongly,goban2015superradiance,lalumiere2013input}.
Recently, experiments have realized multi-point coupling between superconducting
qubits and short-wavelength one-dimensional waveguides, such as surface
acoustic waves (SAWs) devices \citep{Gustafsson2014,Andersson2020,Guo2017,Andersson2019,Wang2022},
microwave transmission lines (TLs) \citep{Astafiev2010,Loo2013,Liu2017},
coupled-resonator waveguides \citep{zheng2025single,wang2024controlling,zheng2010waveguide},
and photonic crystal waveguides \citep{Wang2021,Lim2023}. In these
systems, the usual dipole approximation no longer holds, and such
atom--waveguide configurations are referred to as giant atoms. Self-interference
of giant atoms gives rise to a range of unconventional phenomena,
including frequency-dependent Lamb shifts and decay rates \citep{FriskKockum2014,Manenti2017,Gustafsson2014,Kannan2020,Kockum2021,Vadiraj2021,Soro2022},
non-Markovian decay dynamics \citep{Guo2017,Andersson2019,Cilluffo2020,Guo2020,yin2022non,Qiu2023},
bound states \citep{Wang2021,Mahmoodian2020,BelloXXXX,Lim2023,jia2024atom,weng2025high,yang2025coherent,sun2025cavity},
decoherence-free interactions \citep{Kockum2018,FornDiaz2017,Andersson2019,Carollo2020,Du2023a,Du2023},
and entanglement dynamics \citep{Pompili2021,Sipahigil2016,Yin2023,luo2024entangling,Du2025}.

Giant atoms coupled to waveguides have emerged as a versatile platform
for controlling single-photon transport. By tuning atom-waveguide coupling strengths, distances between coupling points, and atomic
spontaneous emission rates, the spectral response and the distribution
of single photons among output ports can be precisely engineered \citep{Cai2021,Chen2022,Du2021a,gu2024two,Zhao2022,zhu2025single,Feng2021,Peng2023,li2024single}.
When interatomic coupling is introduced, giant atoms can form giant
molecules, which encode atomic interactions directly into the photon\nobreakdash-scattering
behavior \citep{Kannan2023,Chen2014,Almanakly2025,Yin2022,Zhou2023,bao2024efficient,zheng2024chiral,sun2025phase,Gong2024}.
This architecture gives rise to several striking features, including
spectral splitting \citep{Kannan2023,Chen2014,Almanakly2025}, electromagnetically
induced transparency \citep{Yin2022,Zhou2023}, and the ability to
realize single\nobreakdash-photon isolation and directional transmission
\citep{bao2024efficient,sun2025phase,Gong2024,zheng2024chiral}. Despite these
advances, research on giant-molecule systems remains limited---most studies have focused on weak coherent exchanging interactions \citep{Kannan2023,Almanakly2025} or  resonant coupling \citep{Yin2022,Zhou2023}. When the atoms are frequency-detuned, the exchange of excitations becomes retarded, effectively modulating the competition between atom-waveguide coupling and direct interatomic interaction. This modulated competition can strongly influence the photon scattering spectra, potentially giving rise to non-Markovian effects that have yet to be explored.

We present a theoretical study of single\nobreakdash-photon scattering
in a waveguide\nobreakdash-QED system where a giant molecule---composed
of two frequency\nobreakdash-detuned giant atoms with coherent coupling---is
asymmetrically coupled to two parallel waveguides via multiple connection
points. This design provides independent control over the atom--waveguide
couplings, including Lamb shifts and effective decay rates. We develop
a unified framework describing the competition between coherent interatomic
interaction and radiative decay, which classifies the dynamics into
weak, critical, and strong coupling regimes. Interestingly, we show that the optimal scattering probabilities between waveguides emerge from path interference jointly governed by the frequency detuning and the asymmetric decay rates of the giant atoms. Furthermore, photon scattering can be engineered to be directional through the implementation of chiral atom-waveguide couplings. Extending our analysis to the non-Markovian regime reveals that atomic and probe detunings can cooperatively drive transitions between weak- and strong-coupling regimes. The associated retardation effect reshapes the phase diagram into a multiple-resonance landscape, thereby generating a rich structure of peaks and avoided crossings. These results
provide a comprehensive picture of how detuning, coherent coupling,
and retardation feedback cooperate to govern single\nobreakdash-photon
transport in giant\nobreakdash-molecule waveguide\nobreakdash-QED
systems.

The paper is organized as follows. In Section \ref{sec:2}, we introduce
the theoretical model of a giant molecule coupled to two parallel
waveguides. Using a real-space scattering approach, we derive analytical
expressions for the transmission and reflection coefficients of an
incident photon, and discuss the resulting Lamb shifts and effective
decay rates of the giant atoms. Section \ref{sec:3} presents a systematic
analysis of the scattering spectra in the Markovian limit, focusing
on the characteristic features that emerge in the weak, critical,
and strong coupling regimes due to inter\nobreakdash-atomic coupling.
Section \ref{sec:4} extends the study to the non\nobreakdash-Markovian
regime, where time\nobreakdash-delay effects are shown to reshape
the spectra and induce transitions between the weak and strong coupling
regimes. Section \ref{sec:5} investigates the scattering properties
for perfect chiral\nobreakdash-coupling in the Markovian and non-Markovian
regimes, and discusses the experimental feasibility of the system.
Finally, Section \ref{sec:6} provides a summary of the main results
and concludes the paper.

\section{MODEL AND METHOD\label{sec:2}}

\begin{figure}
\includegraphics[width=1\columnwidth]{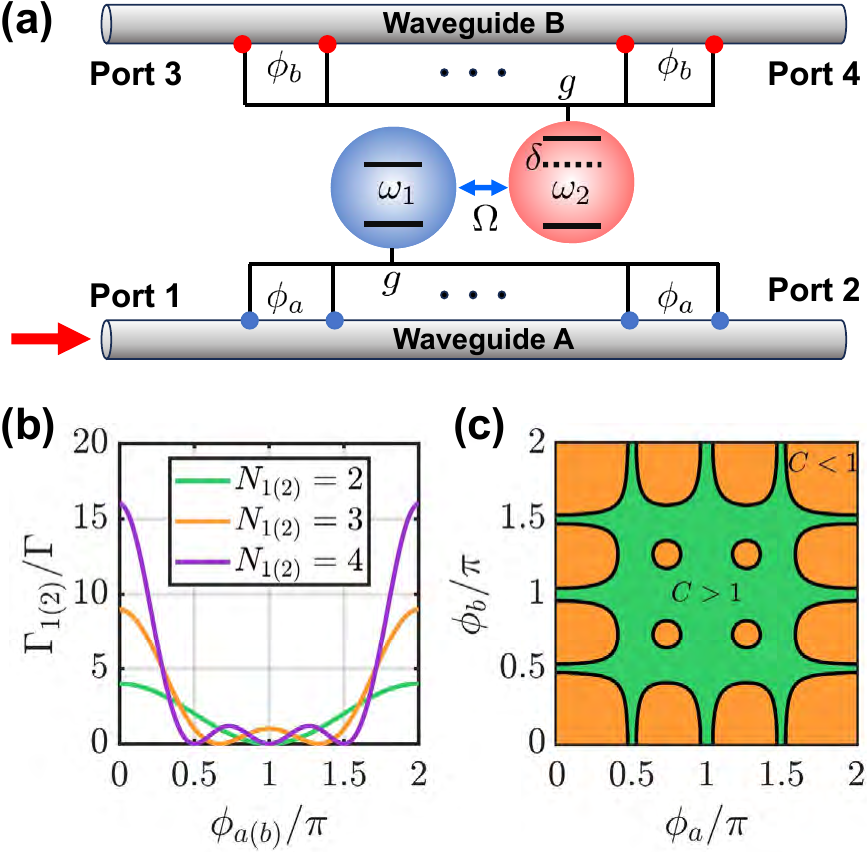}\caption{\label{Fig.1Model}(a) Schematics of a giant-molecule waveguide-QED
setup, where two interacting giant atoms independently couple to two
parallel waveguides via multiple connection points. (b) Effective
decay rates $\Gamma_{1(2)}$ as a function of the phase $\phi_{a(b)}$
for different numbers of coupling points $N_{1(2)}=2,3,4$. (c) Phase
diagram of the cooperativity parameter\textcolor{red}{{} }$C$ in the
$(\phi_{a},\phi_{b})$ plane for $N_{1}=N_{2}=4$. The interatomic
coupling strength is set to $\Omega/\Gamma=1$. The green (orange)
region corresponds to $C>1$ ($C<1$).}
\end{figure}

As sketched in Fig. \ref{Fig.1Model}(a), we investigate a giant molecule
comprising two giant atoms with distinct transition frequencies $\omega_{1}$
and $\omega_{2}$, which are subject to a coupler-mediated exchange
interaction (with strength $\Omega$). The two giant atoms are asymmetrically
coupled to parallel waveguides A and B through multiple discrete connection
points. Atom\,1 connects to waveguide\,A at positions $\{z_{j}^{a}\}_{j=1}^{N_{1}}$,
and atom\,2 to waveguide\,B at positions $\{z_{j}^{b}\}_{j=1}^{N_{2}}$,
with $N_{1(2)}\geqslant2$. Within each waveguide, the coupling points
are equally spaced, but the spacing intervals can differ between waveguide\,A
and waveguide\,B. For simplicity, we assume a uniform coupling strength
$g$ at every connection point. In the rotating-wave approximation
(RWA), and setting $\hbar=1$, the total Hamiltonian of the system
in real space takes the form \citep{Kannan2023,Almanakly2025}:
\begin{align}
H & =H_{gm}+H_{w}+H_{int}\label{eq:Hamiltion}
\end{align}
with
\begin{equation}
H_{gm}= \sum_{j=1,2}\omega_{j}\hat{\sigma}_{j}^{+}\hat{\sigma}_{j}^{-}+\Omega(\hat{\sigma}_{1}^{-}\hat{\sigma}_{2}^{+}+\hat{\sigma}_{1}^{+}\hat{\sigma}_{2}^{-}), \notag
\end{equation}

\begin{equation}
H_{w}= \sum_{l=L,R}if_{l}v_{g}\int_{-\infty}^{+\infty}dz\left[a_{l}^{\dagger}(z)\frac{\partial}{\partial z}a_{l}(z)+b_{l}^{\dagger}(z)\frac{\partial}{\partial z}b_{l}(z)\right], \notag
\end{equation}

\begin{equation}
\begin{split}
H_{int}= & g\int_{-\infty}^{+\infty}dz\sum_{l=L,R}\sum_{j=1}^{N_{1}}\delta(z-z_{j}^{a})a_{l}^{\dagger}(z)\hat{\sigma}_{1}^{-} \\
 & + g\int_{-\infty}^{+\infty}dz\sum_{l=L,R}\sum_{j=1}^{N_{2}}\delta(z-z_{j}^{b})b_{l}^{\dagger}(z)\hat{\sigma}_{2}^{-} + \text{H.c.},
\end{split} \notag
\end{equation}
where $H_{gm}$ is the bare Hamiltonian for the giant molecule, $\hat{\sigma}_{1}^{+}=|e\rangle_{1}\langle g|$
($\hat{\sigma}_{2}^{+}=|e\rangle_{2}\langle g|$) and $\hat{\sigma}_{1}^{-}=(\hat{\sigma}_{1}^{+})^{\dagger}$
{[}$\hat{\sigma}_{2}^{-}=(\hat{\sigma}_{2}^{+})^{\dagger}${]} are
atomic raising and lowering operators with $|g\rangle_{1(2)}$ and
$|e\rangle_{1(2)}$ the ground and excited states of atom 1 (2). The
term $H_{w}$ describes the free propagation of photons in the two
waveguides. Here, $f_{L}=1$ and $f_{R}=-1$ indicate the left- and
right-propagation directions, respectively, while $a_{L}(z)$ and
$a_{R}(z)$ {[}$b_{L}(z)$ and $b_{R}(z)${]} denote the annihilation
operators for left- and right-propagating photons at position $z$
in waveguide A (B). We consider a linear dispersion relation $\omega(k)=kv_{g}$,
with $k$ the wave vector \textbf{(taken to be positive without loss of generality)}, and $v_{g}$ the group velocity for the
two linear waveguides. The interaction Hamiltonian $H_{int}$, which
governs the coupling between the giant molecule and the waveguide
A (B) with $N_{1}$ ($N_{2}$) connection points, is implemented in
real space through localized delta functions $\delta(z-z_{j}^{a(b)})$
at each connection point. A striking feature of this architecture
is that the system can exhibit either Markovian or non-Markovian dynamics,
controlled by the accumulated phase $\phi_{a(b)}=k|z_{j+1}^{a(b)}-z_{j}^{a(b)}|$
acquired by a photon propagating between neighboring coupling points.

We investigate the transport of a photon (with frequency $\omega(k)$
and wave vector $k>0$) incident from waveguide A and scattered by
the giant molecule. In particular, we focus on the regime where the
transition frequency difference between the two atoms, characterized
by the detuning $\delta=\omega_{2}-\omega_{1}$, plays a significant
role. This detuning introduces retardation in the excitation exchange,
setting up a competition between radiative decay into the waveguides
and coherent atom--atom coupling. Consequently, $\delta$ can dramatically
alter the photon scattering probabilities and generate non-Markovian
time delays between the waveguides. In the single-excitation subspace,
the eigenstates of the total Hamiltonian (\ref{eq:Hamiltion}) can
be expressed in the general form:
\begin{align}
\left|\Psi\right\rangle  & =\underset{l=L,R}{\sum}\int_{-\infty}^{+\infty}dz[\psi_{l}^{a}(z)a_{l}^{\dagger}(z)+\psi_{l}^{b}(z)b_{l}^{\dagger}(z)]\left|0,g_{1},g_{2}\right\rangle \nonumber \\
 & +u_{1}\left|0,e_{1},g_{2}\right\rangle +u_{2}\left|0,g_{1},e_{2}\right\rangle ,\label{eq:eigenstate}
\end{align}
where $\left|0,g_{1},g_{2}\right\rangle $ represents the vacuum state
of both waveguides with the two atoms in their ground states. $\psi_{l}^{a}(z)$
and $\psi_{l}^{b}(z)$ (with $l=L,R$) are the wave functions for
left- and right-propagating photons in waveguides A and B, respectively.
The coefficients $u_{1}$ and $u_{2}$ are the probability amplitudes
for atoms $1$ and $2$ to be in the excited states $|e_{1}\rangle$
and $|e_{2}\rangle$, with no photons present in either waveguide.
Inserting Eq. (\ref{eq:eigenstate}) into the stationary Schr\"odinger
equation $H\left|\Psi\right\rangle =E\left|\Psi\right\rangle $ leads
to a set of equations for the probability amplitudes:
\begin{equation}
E\psi_{L}^{a}(z) = iv_{g}\frac{\partial}{\partial z}\psi_{L}^{a}(z)+g\sum_{j=1}^{N_{1}}\delta(z-z_{j}^{a})u_{1}, \notag
\end{equation}
\begin{equation}
E\psi_{R}^{a}(z) = -iv_{g}\frac{\partial}{\partial z}\psi_{R}^{a}(z)+g\sum_{j=1}^{N_{1}}\delta(z-z_{j}^{a})u_{1}, \notag
\end{equation}
\begin{equation}
E\psi_{L}^{b}(z) = iv_{g}\frac{\partial}{\partial z}\psi_{L}^{b}(z)+g\sum_{j=1}^{N_{2}}\delta(z-z_{j}^{b})u_{2}, \notag
\end{equation}
\begin{equation}
E\psi_{R}^{b}(z) = -iv_{g}\frac{\partial}{\partial z}\psi_{R}^{b}(z)+g\sum_{j=1}^{N_{2}}\delta(z-z_{j}^{b})u_{2}, \notag
\end{equation}
\begin{equation}
\Delta u_{1} = g\sum_{j=1}^{N_{1}}\delta(z-z_{j}^{a})\bigl[\psi_{L}^{a}(z)+\psi_{R}^{a}(z)\bigr]+\Omega u_{2}, \notag
\end{equation}
\begin{equation}
(\Delta-\delta)u_{2} = g\sum_{j=1}^{N_{2}}\delta(z-z_{j}^{b})\bigl[\psi_{L}^{b}(z)+\psi_{R}^{b}(z)\bigr]+\Omega u_{1},
\end{equation}
where $\Delta=\omega(k)-\omega_{1}$ and $\Delta-\delta$ are the
detunings of the incident photon relative to atoms\,1 and\,2, respectively.
Assuming the photon enters from port\,1 (left side of waveguide\,A),
the wavefunctions in the two waveguides take the form:
\begin{eqnarray*}
\psi_{R}^{a}(z) & = & e^{ikz}[\Theta(z_{1}^{a}-z)\\
 &  & +\stackrel[j=1]{N_{1}-1}{\sum}t_{j}^{a}\Theta(z-z_{j}^{a})\Theta(z_{j+1}^{a}-z) \\ &  & + t_{N_{1}}^{a}\Theta(z-z_{N_{1}}^{a})],
\end{eqnarray*}
\begin{eqnarray*}
\psi_{L}^{a}(z) & = & e^{-ikz}[r_{1}^{a}\Theta(z_{1}^{a}-z)\\
 &  & +\stackrel[j=2]{N_{1}-1}{\sum}r_{j}^{a}\Theta(z-z_{j-1}^{a})\Theta(z_{j}^{a}-z)],
\end{eqnarray*}
\begin{eqnarray*}
\psi_{R}^{b}(z) & = & e^{ikz}[\stackrel[j=1]{N_{2}-1}{\sum}t_{j}^{b}\Theta(z-z_{j}^{b})\Theta(z_{j+1}^{b}-z)\\
 &  & +t_{N_{2}}^{b}\Theta(z-z_{N_{2}}^{b})],
\end{eqnarray*}
\begin{eqnarray}
\psi_{L}^{b}(z) & = & e^{-ikz}[r_{1}^{b}\Theta(z_{1}^{b}-z)\nonumber \\
 &  & +\stackrel[j=2]{N_{2}-1}{\sum}r_{j}^{b}\Theta(z-z_{j-1}^{b})\Theta(z_{j}^{b}-z)],\label{eq:LeftAmplitude}
\end{eqnarray}
where $t_{j}^{a(b)}$ and $r_{j}^{a(b)}$ are the transmission and
reflection coefficients at the $j$-th coupling point in waveguide
A (B) with $\Theta(z-z_{j}^{a(b)})$ the Heaviside step function.
The scattering amplitudes for a photon incident from port\,1 are defined
as follows. The reflection amplitude back into port\,1 is given by
$r_{1\rightarrow1}\equiv r_{1}^{a}$, and the transmission amplitude
into port\,2 by $t_{1\rightarrow2}\equiv t_{N_{1}}^{a}$; the amplitudes
for transfer into waveguide\,B are defined as $t_{1\rightarrow3}\equiv r_{1}^{b}$
(transmission to port\,3) and $t_{1\rightarrow4}\equiv t_{N_{2}}^{b}$
(transmission to port\,4). The corresponding transmission and reflection
probabilities are $T_{1\rightarrow2}=|t_{1\rightarrow2}|^{2}$, $R_{1\rightarrow1}=|r_{1\rightarrow1}|^{2}$
and $T_{1\rightarrow3(4)}=|t_{1\rightarrow3(4)}|^{2}$, which satisfy
the probability-conservation (normalization) condition: $T_{1\rightarrow2}+R_{1\rightarrow1}+T_{1\rightarrow4}+T_{1\rightarrow3}=1$.
Substituting the ansatz from Eqs.\,(\ref{eq:LeftAmplitude}) into
the stationary Schr\"odinger equation yields explicit expressions for
the scattering amplitudes \citep{li2024single}:
\begin{align}
t_{1\rightarrow2} & =\frac{\Omega^{2}-\widetilde{\Delta}[(\widetilde{\Delta}-\widetilde{\delta})+i\Gamma_{2}]}{\Omega^{2}-[\widetilde{\Delta}+i\Gamma_{1}][(\widetilde{\Delta}-\widetilde{\delta})+i\Gamma_{2}]},\nonumber \\
r_{1\rightarrow1} & =\frac{i\Gamma_{1}[(\widetilde{\Delta}-\tilde{\delta})+i\Gamma_{2}]}{\Omega^{2}-[\widetilde{\Delta}+i\Gamma_{1}][(\widetilde{\Delta}-\widetilde{\delta})+i\Gamma_{2}]},\nonumber \\
t_{1\rightarrow3(4)} & =\frac{i\sqrt{\Gamma_{1}\Gamma_{2}}\Omega}{\Omega^{2}-[\widetilde{\Delta}+i\Gamma_{1}][(\widetilde{\Delta}-\widetilde{\delta})+i\Gamma_{2}]},\label{eq:t and r-1}
\end{align}
where $\widetilde{\Delta}=\Delta-\Delta_{\text{ls},1}$ and $\widetilde{\delta}=\delta-\Delta_{\text{ls},1}+\Delta_{\text{ls},2}$.
The overall phase factors have been omitted. The Lamb shifts $\Delta_{\text{ls},1}$
and $\Delta_{\text{ls},2}$ induced by the atom-waveguide coupling
are given by
\begin{align}
\Delta_{\text{ls},1(2)} & =\Gamma\frac{N_{1(2)}\sin\phi_{a(b)}-\sin[N_{1(2)}\phi_{a(b)}]}{1-\cos\phi_{a(b)}}.\label{eq:Delta_lsa}
\end{align}
The corresponding effective decay rates of atom 1 (2) into the waveguide
A (B) are
\begin{align}
\Gamma_{1(2)} & =\Gamma\frac{\sin^{2}[\frac{1}{2}N_{1(2)}\phi_{a(b)}]}{\sin^{2}[\frac{1}{2}\phi_{a(b)}]},\label{eq:Gamma_a}
\end{align}
with $\Gamma=g^{2}/v_{g}$ is the radiative decay rate for a single
coupling point (i.e., the small\nobreakdash-atom limit). Both the
Lamb shifts and the effective decay rates depend on the number of
coupling points and the spacing between them.

In Fig.\, \ref{Fig.1Model}(b), we plot $\Gamma_{1}$ and $\Gamma_{2}$
as functions of $\phi_{a}$ and $\phi_{b}$, respectively, for $N_{1(2)}=2,3,4$.
By choosing different numbers of coupling points or different coupling-point
separations in the two waveguides, one can make $\Gamma_{1}$ and
$\Gamma_{2}$ differ substantially. As shown below, this tunability
is essential for controlling photon scattering between the waveguides.
There are two special cases in which the giant molecule decouples
from one of the waveguides: from waveguide A when $\Gamma_{1}=0$,
and from waveguide B when $\Gamma_{2}=0$. When $\Gamma_{1}=0$, the
system reduces to a trivial case with perfect transmission: $|t_{1\rightarrow2}|=1$
and $r_{1\rightarrow1}=t_{1\rightarrow3(4)}=0$. For $\Gamma_{2}=0$,
the transmission amplitude $t_{1\rightarrow3(4)}=0$ and $t_{1\rightarrow2}$
simplifies to
\begin{align}
t_{1\rightarrow2} & =\frac{\Omega^{2}-\widetilde{\Delta}(\widetilde{\Delta}-\widetilde{\delta})}{\Omega^{2}-(\widetilde{\Delta}+i\Gamma_{1})(\widetilde{\Delta}-\widetilde{\delta})},\label{eq: EIT}
\end{align}
which is algebraically equivalent to the standard electromagnetically
induced transparency (EIT) transmission profile \citep{Feng2021}.
The transparency windows, where $|t_{1\rightarrow2}|=1$, are centered
at $\Delta=\Delta_{\text{ls},2}+\delta$ (i.e., when $\widetilde{\Delta}=\widetilde{\delta}$).
A strong interatomic coupling $\Omega$ then leads to anti-crossings
in the transmission and reflection spectra---a signature of broken
symmetries. The frequencies of the resulting peaks (or dips) are given
by the real roots of the numerator: $\Delta_{\pm}=[\Delta_{\text{ls},1}+\Delta_{\text{ls},2}+\delta\pm\sqrt{(\Delta_{\text{ls},1}-\Delta_{\text{ls},2}-\delta)^{2}+4\Omega^{2}}]/2$.
Note that if atom\,2 is coupled to waveguide B via only two points,
achieving $\Gamma_{2}=0$ requires setting $\phi_{b}=\pi$, which
simultaneously gives $\Delta_{\text{ls},2}=0$. Consequently, the
transparency window is fixed at $\widetilde{\Delta}=\widetilde{\delta}$.
In contrast, with more than two coupling points, one can achieve $\Gamma_{2}=0$
while maintaining a finite Lamb shift $\Delta_{\text{ls},2}$, thereby
enabling the transparency window to be tuned by $\Delta_{\text{ls},2}$.

In this work, we focus on the general case where both decay rates
are non-vanishing: $\Gamma_{1}\neq0$ and $\Gamma_{2}\neq0$. Moreover,
we introduce the cooperativity parameter $C=\Omega^{2}/(\Gamma_{1}\Gamma_{2})$,
which quantifies the relative strength of coherent atom--atom coupling
to dissipation into the waveguides, and thus serves as a dimensionless
measure of the system's ability to sustain coherent
quantum effects in the presence of engineered dissipation. Based on
the value of the cooperativity parameter $C$, the system can be classified
into three regimes: weak ($C<1$), critical ($C=1$), and strong ($C>1$)
coupling. In Fig. \ref{Fig.1Model}(c), we plot the cooperativity
parameter $C$ as a function of the propagation phases $\phi_{a}$
and $\phi_{b}$ for the equal number of coupling points $N_{1}=N_{2}=4$
and fixed exchange interaction strength $\Omega/\Gamma=1$. These
phases determine the coupling regime in which the system operates.
For each regime---weak, critical, and strong coupling---we then
examine how the scattering of an incident photon is governed by two
key detunings: the frequency detuning $\delta$ between the two atoms
and the probe detuning $\Delta$ relative to atom 1.

\section{photon scattering by a giant molecule: The Markovian regime\label{sec:3}}

We first examine the Markovian regime, where the photon propagation
time between the outermost coupling points in each waveguide, given
by $N_{1(2)}\tau^{a(b)}$ with $\tau^{a(b)}=|z_{2}^{a(b)}-z_{1}^{a(b)}|/v_{g}$,
is negligible compared to the characteristic relaxation time of the
giant atom 1 (2), $\sim1/\left[N_{1(2)}^{2}\Gamma\right]$. Under
this condition, the time delay associated with the multi\nobreakdash-point
coupling can be safely neglected. Specifically, using a linear dispersion
relation around the central frequencies $\omega_{1}$ and $\omega_{2}$,
the photon propagation phases between neighboring coupling points
in waveguides A and B can be expressed as $\phi_{a}=\Delta\tau^{a}+\widetilde{\phi}_{a}$
and $\phi_{b}=\left(\Delta-\delta\right)\tau^{b}+\widetilde{\phi}_{b}$,
where $\widetilde{\phi}_{a}=k_{1}|z_{2}^{a}-z_{1}^{a}|$ (with $k_{1}=\omega_{1}/v_{g}$)
and $\widetilde{\phi}_{b}=k_{2}|z_{2}^{b}-z_{1}^{b}|$ (with $k_{2}=\omega_{2}/v_{g}$).
For a finite bandwidth on the order of $N_{1(2)}^{2}\Gamma$, the
Markovian condition requires $\{|\Delta|\tau^{a},|\Delta-\delta|\tau^{b}\}\ll1$.
Under these conditions, we can approximate $\phi_{a(b)}\approx\widetilde{\phi}_{a(b)}$,
which simplifies the evaluation of the Lamb shifts and effective decay
rates in Eqs. (\ref{eq:Delta_lsa}) and (\ref{eq:Gamma_a}).

To gain physical insight into the scattering behavior, we first analyze
the condition $\widetilde{\Delta}=0$ in which the incident photon
is on resonance with atom\,1. In this case, the scattering amplitudes
take the simpler form:
\[
t_{1\rightarrow2}(\widetilde{\delta})=\frac{C}{C+1+i\tilde{\delta}\Gamma_{2}^{-1}},\text{ }r_{1\rightarrow1}(\widetilde{\delta})=\frac{-1-i\tilde{\delta}\Gamma_{2}^{-1}}{C+1+i\tilde{\delta}\Gamma_{2}^{-1}},
\]
\begin{eqnarray}
t_{1\rightarrow3(4)} & (\widetilde{\delta})= & \frac{i\sqrt{C}}{C+1+i\tilde{\delta}\Gamma_{2}^{-1}}.\label{eq:T(Delta=00003DDelta_ls,a)}
\end{eqnarray}
Both $t_{1\rightarrow2}$ and $t_{1\rightarrow3(4)}$ are symmetric
Lorentzian in $\widetilde{\delta}$, centered at $\widetilde{\delta}=0$
with a full width at half maximum (FWHM) of $2(C+1)\Gamma_{2}$. The
resonant transfer probability to waveguide\,B, given by $|t_{1\rightarrow3(4)}(0)|^{2}=C/(C+1)^{2}$,
reaches its maximum value of 0.25 when $C=1$. For $C>1$, increasing
$C$ broadens the resonance linewidth while reducing the peak amplitude.

Second, when the incident photon is on resonance with atom 2, i.e.
$\widetilde{\Delta}=\widetilde{\delta}$ or equivalently $\Delta-\delta=\Delta_{\text{ls},2}$,
the scattering amplitudes become
\[
t_{1\rightarrow2}(\widetilde{\delta})=1-\frac{1}{C+1-i\tilde{\delta}\Gamma_{1}^{-1}},\text{ }r_{1\rightarrow1}(\widetilde{\delta})=\frac{-1}{C+1-i\tilde{\delta}\Gamma_{1}^{-1}},
\]
\begin{eqnarray}
t_{1\rightarrow3(4)} & (\widetilde{\delta})= & \frac{i\sqrt{C}}{C+1-i\tilde{\delta}\Gamma_{1}^{-1}}.\label{eq:T(Delta=00003DDelta_ls,b)}
\end{eqnarray}
The resonant transfer probability to waveguide\,B, $|t_{1\rightarrow3(4)}(0)|=C/(C+1)^{2}$,
again reaches a maximum of 0.25 at $C=1$; but the FWHM of the resonance
is now $2(C+1)\Gamma_{1}$. Defining the ratio $\mathcal{R}\equiv\text{\ensuremath{\Gamma_{1}}/\ensuremath{\Gamma_{2}}}$,
this width can also be written as $2(C+1)\Gamma_{2}\mathcal{R}$.
Compared with the case of resonance with atom\,1 ($\tilde{\Delta}=0$),
where the FWHM was $2(C+1)\Gamma_{2}$, the linewidth is therefore
scaled by the factor $\mathcal{R}$. \textit{However, for $C>1$,
photon transfer to waveguide\,B is not maximized under atomic detuning
($\delta\neq0$), neither at the resonant condition $\widetilde{\Delta}=0$
nor at $\widetilde{\Delta}=\widetilde{\delta}$.}

\subsection{Weak coupling regime: $C\ll1$}

\begin{figure}
\includegraphics[width=1\columnwidth]{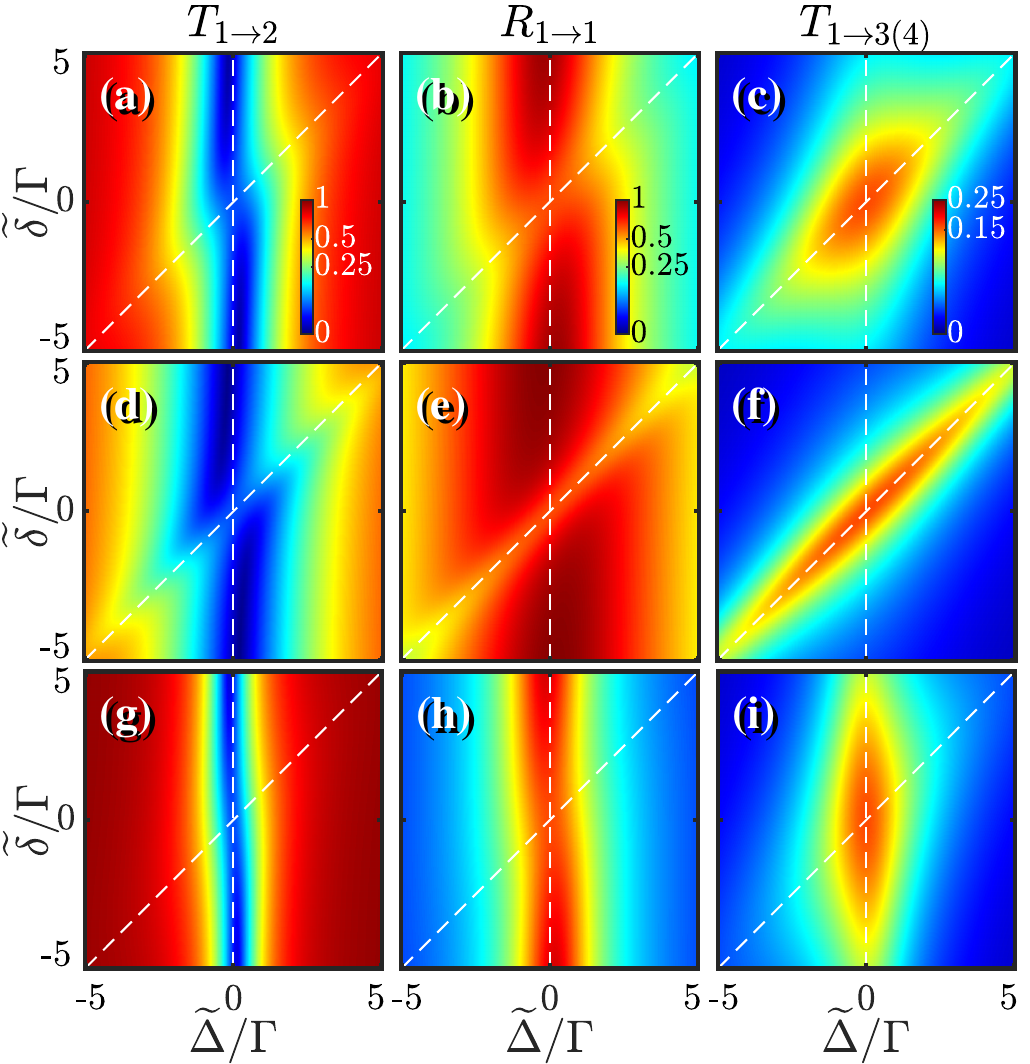}

\caption{\label{Fig.2Weak}\textbf{ The weak coupling regime.} Scattering probabilities
$T_{1\rightarrow2}$, $R_{1\rightarrow1}$ and $T_{1\rightarrow3(4)}$
as functions of $\widetilde{\Delta}/\Gamma$ and $\widetilde{\delta}/\Gamma$
with the cooperativity parameter $C=1/4$ and the number of coupling
points $N_{1}=N_{2}=4$. The exchange interatomic coupling strength
is set to $\Omega/\Gamma=1$. The effective decay rates are $(\Gamma_{1},\Gamma_{2})/\Omega=(2,2)$
with $(\widetilde{\phi}_{a},\widetilde{\phi}_{b})/\pi\simeq(9/25,9/25)$
{[}panels (a)--(c){]}, $(\Gamma_{1},\Gamma_{2})/\Omega=(4,1)$ with
$(\widetilde{\phi}_{a},\widetilde{\phi}_{b})/\pi\simeq(4/13,4/5)$
{[}panels (d)--(f){]}, and $(\Gamma_{1},\Gamma_{2})/\Omega=(1,4)$
with $(\widetilde{\phi}_{a},\widetilde{\phi}_{b})/\pi\simeq(4/5,4/13)$
{[}panels (g)--(i){]}. The white dashed lines indicate $\widetilde{\Delta}=0$
and $\widetilde{\Delta}=\widetilde{\delta}$, respectively.}
\end{figure}

In Fig. \ref{Fig.2Weak}, we show the scattering probabilities for
finding the photon at the four ports as functions of $\widetilde{\Delta}/\Gamma$
and $\widetilde{\delta}/\Gamma$ with $(\Gamma_{1},\Gamma_{2})/\Omega=(2,2)$
{[}$(\widetilde{\phi}_{a},\widetilde{\phi}_{b})/\pi\simeq(9/25,9/25)$,
Figs. \ref{Fig.2Weak}(a)--\ref{Fig.2Weak}(c){]}, $(\Gamma_{1},\Gamma_{2})/\Omega\simeq(4,1)$
{[}$(\widetilde{\phi}_{a},\widetilde{\phi}_{b})/\pi\simeq(4/13,4/5)$,
Figs. \ref{Fig.2Weak}(d)--\ref{Fig.2Weak}(f){]}, $(\Gamma_{1},\Gamma_{2})/\Omega=(1,4)$
{[}$(\widetilde{\phi}_{a},\widetilde{\phi}_{b})/\pi\simeq(4/5,4/13)$,
Figs. \ref{Fig.2Weak}(g)--\ref{Fig.2Weak}(i){]}, respectively.
In all cases, the cooperativity parameter is $C=1/4$. As shown in
Figs. \ref{Fig.2Weak}(a)--\ref{Fig.2Weak}(c) for $C\ll1,$ the
transmission to port 2 $T_{1\rightarrow2}(\widetilde{\Delta})$ {[}or
the reflection $R_{1\rightarrow1}(\widetilde{\Delta})${]} exhibit
an asymmetric anti-Lorentzian (or Lorentzian) lineshape for fixed
$|\widetilde{\delta}|$. The minimum of $T_{1\rightarrow2}(\widetilde{\Delta})$
{[}and the maximum of $R_{1\rightarrow1}(\widetilde{\Delta})${]}
occurs at $\widetilde{\Delta}=0$. Along the white dashed line ($\widetilde{\Delta}=0$),
increasing the interatomic detuning $|\widetilde{\delta}|$ suppresses
the transmission as $T_{1\rightarrow2}(\widetilde{\Delta}=0,\widetilde{\delta})\sim\frac{C^{2}}{1+(\widetilde{\delta}/\Gamma_{b})^{2}}$,
while increasing $R_{1\rightarrow1}(\widetilde{\Delta}=0,\widetilde{\delta})\rightarrow1$.
In this limit, atom\,2 is far off\nobreakdash-resonance and cannot
be efficiently excited, so the system reduces to an effective single\nobreakdash-atom
scattering process where atom\,1 acts as a mirror. The transfer probabilities
to waveguide\,B, $T_{1\rightarrow3(4)}(\widetilde{\Delta},\widetilde{\delta})$,
reach their peak values $\sim C/(C+1)^{2}$ at $\widetilde{\Delta}=\widetilde{\delta}=0$.

Notably, the ratio $\mathcal{R}$, which quantifies the asymmetry
in the effective decay rates, can strongly modify the scattering spectra.
For $\mathcal{R}>1$ {[}see Figs. \ref{Fig.2Weak}(d)--\ref{Fig.2Weak}(f){]},
the scattering profiles are stretched along the line $\widetilde{\Delta}=\widetilde{\delta}$,
producing weakly resolved double dips (peaks) in the transmission
(reflection) spectra $T_{1\rightarrow2}(\widetilde{\Delta})$ {[}$R_{1\rightarrow1}(\widetilde{\Delta})${]}.
Conversely, for $\mathcal{R}<1$ {[}see Figs. \ref{Fig.2Weak}(g)--\ref{Fig.2Weak}(i){]},
the scattering features become concentrated along the line $\widetilde{\Delta}=0$.
For both $\mathcal{R}>1$ and $\mathcal{R}<1$, the peak values of
$T_{1\rightarrow3(4)}$ are consistently found at $\widetilde{\Delta}=\widetilde{\delta}=0$
and are well approximated by $C/(C+1)^{2}$ (with $C=1/4$ here).

\subsection{Critical coupling regime: $C\sim1$}

\begin{figure}
\includegraphics[width=1\columnwidth]{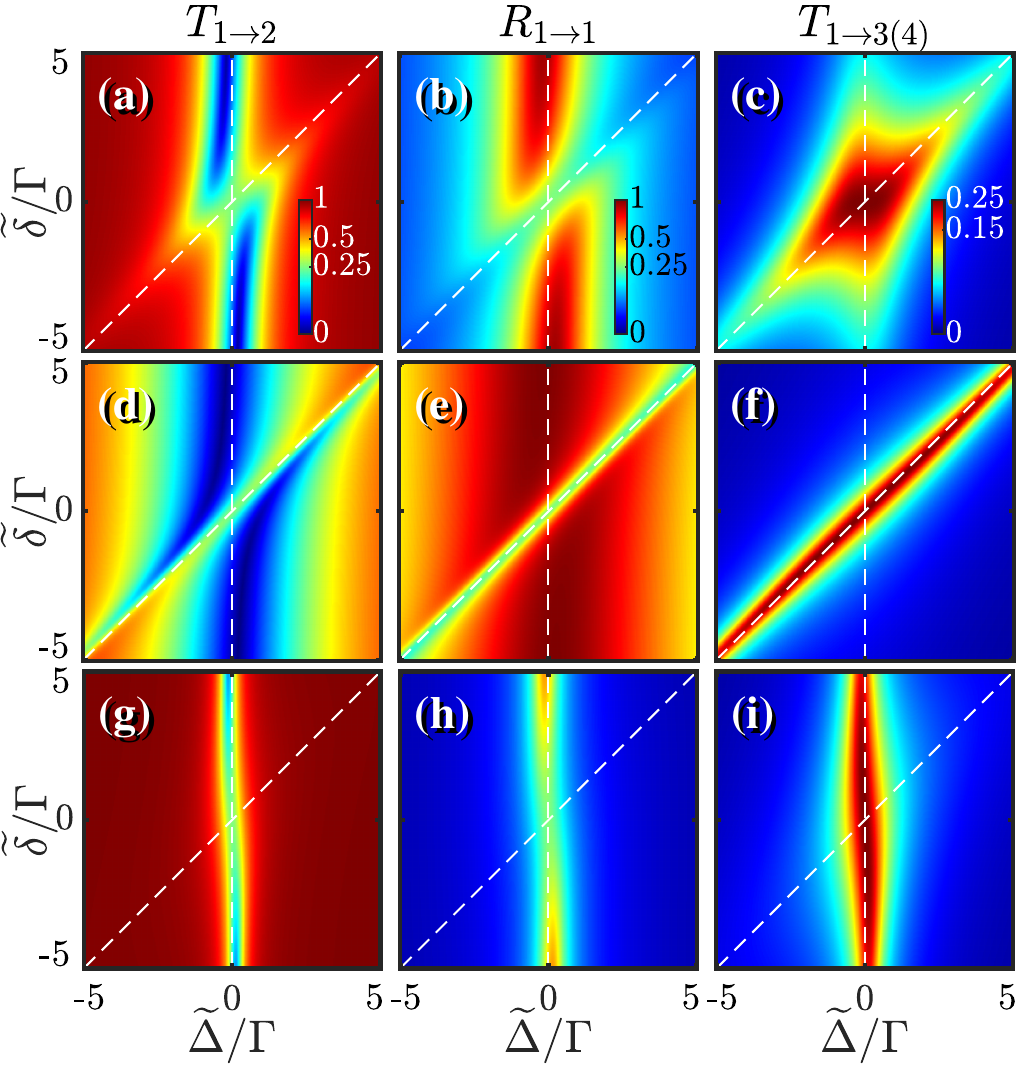}

\caption{\label{Fig.3Intermedia}\textbf{The critical coupling regime.} Scattering
probabilities $T_{1\rightarrow2}$, $R_{1\rightarrow1}$ and $T_{1\rightarrow3(4)}$
as functions of $\widetilde{\Delta}/\Gamma$ and $\widetilde{\delta}/\Gamma$
for $C=1$, $N_{1}=N_{2}=4$, and $\Omega/\Gamma=1$. The effective
decay rates are $(\Gamma_{1},\Gamma_{2})/\Omega=(1,1)$ with $(\widetilde{\phi}_{a},\widetilde{\phi}_{b})/\pi\simeq(4/5,4/5)$
{[}panels (a)--(c){]}, $(\Gamma_{1},\Gamma_{2})/\Omega=(4,1/4)$
with $(\widetilde{\phi}_{a},\widetilde{\phi}_{b})/\pi\simeq(4/13,9/16)$
{[}panels (d)--(f){]}, and $(\Gamma_{1},\Gamma_{2})/\Omega=(1/4,4)$
with $(\widetilde{\phi}_{a},\widetilde{\phi}_{b})/\pi\simeq(9/16,4/13)$
{[}panels (g)--(i){]}. The white dashed lines indicate $\widetilde{\Delta}=0$
and $\widetilde{\Delta}=\widetilde{\delta}$, respectively.}
\end{figure}

In the critical coupling regime, we consider three representative
cases with with $C=1$: $(\Gamma_{1},\Gamma_{2})/\Omega=(1,1)$ {[}$(\widetilde{\phi}_{a},\widetilde{\phi}_{b})/\pi\simeq(4/5,4/5)$,
Figs. \ref{Fig.3Intermedia}(a)--\ref{Fig.3Intermedia}(c){]}, $(\Gamma_{1},\Gamma_{2})/\Omega=(4,1/4)$
{[}$(\widetilde{\phi}_{a},\widetilde{\phi}_{b})/\pi\simeq(4/13,9/16)$,
Figs. \ref{Fig.3Intermedia}(d)--\ref{Fig.3Intermedia}(f){]}, $(\Gamma_{1},\Gamma_{2})/\Omega=(1/4,4)$
{[}$(\widetilde{\phi}_{a},\widetilde{\phi}_{b})/\pi\simeq(9/16,4/13)$
{[}Figs. \ref{Fig.3Intermedia}(g)--\ref{Fig.3Intermedia}(i){]}.
These correspond to asymmetry ratios $\mathcal{R}=1$, $\mathcal{R}=16$,
and $\mathcal{R}=1/16$, respectively. For $\mathcal{R}\geq1$, the
two valleys in transmission $T_{1\rightarrow2}$ (and the two peaks
in reflection $R_{1\rightarrow1}$) become increasingly separated
along the white line ($\widetilde{\Delta}=\widetilde{\delta}$) {[}see
Figs. \ref{Fig.3Intermedia}(a), \ref{Fig.3Intermedia}(b), \ref{Fig.3Intermedia}(d),
\ref{Fig.3Intermedia}(e){]}. In particular, the incident photon is
strongly reflected when $\widetilde{\Delta}=0$ and $\widetilde{\delta}/\Gamma\sim5$.
For $\mathcal{R}\ll1$, transmission $T_{1\rightarrow2}$ (and reflection
$R_{1\rightarrow1}$) increases (decreases) sharply along the white
dashed line ($\widetilde{\Delta}=0$) with a narrow width $\sim\Gamma_{1}$
{[}see Figs. \ref{Fig.3Intermedia}(g) and \ref{Fig.3Intermedia}(h){]}.
In all three cases, the photon can be transmitted into waveguide B
with a total probability $T_{1\rightarrow3}+T_{1\rightarrow4}=1/2$
at $\widetilde{\Delta}=\widetilde{\delta}=0$ {[}Figs. \ref{Fig.3Intermedia}(c),
\ref{Fig.3Intermedia}(f), and \ref{Fig.3Intermedia}(i){]}. The ridge
structure of the transfer probabilities $T_{1\rightarrow3(4)}$ is
pronounced along the diagonal $\widetilde{\Delta}=\widetilde{\delta}$
when $\mathcal{R}\gg1$, and along $\widetilde{\Delta}=0$ when $\mathcal{R}\ll1$.
In each case, the $\widetilde{\Delta}$-dependence of $T_{1\rightarrow3(4)}$
are Lorentzian, with a FWHM of $4\Gamma_{1}$ for $\mathcal{R}\ll1$
and $4\Gamma_{2}$ for $\mathcal{R}\gg1$, respectively. Finally,
at $\widetilde{\Delta}=\widetilde{\delta}=0$ with $C=1$, the scattering
probabilities can be evenly partitioned among all four ports: $T_{1\rightarrow2}=R_{1\rightarrow1}=T_{1\rightarrow4}=R_{1\rightarrow3}=1/4$.

\subsection{Strong coupling regime: $C\text{\ensuremath{\gg}}1$}

\begin{figure}
\includegraphics[width=1\columnwidth]{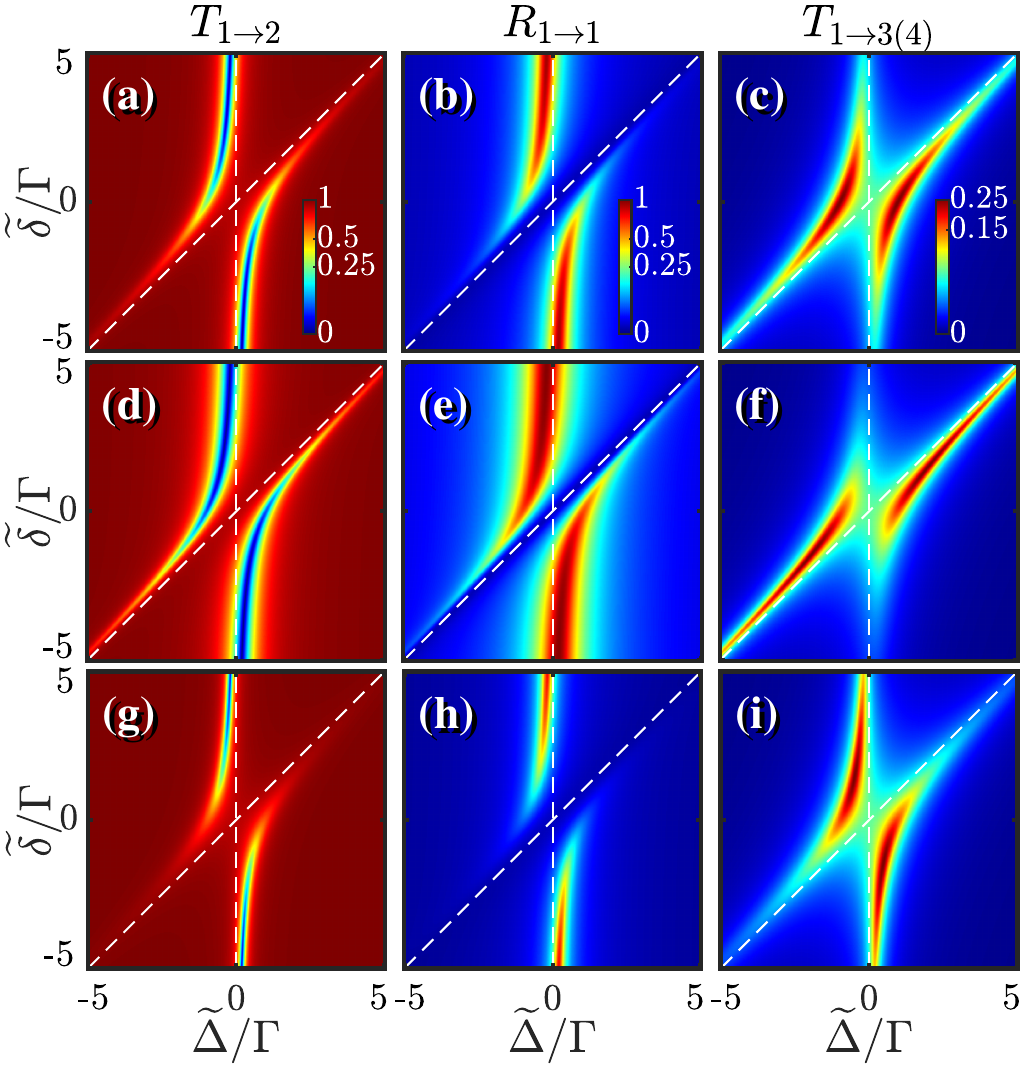}

\caption{\label{Fig.4 Strong}\textbf{The strong coupling regime.} Scattering
probabilities $T_{1\rightarrow2}$, $R_{1\rightarrow1}$ and $T_{1\rightarrow3(4)}$
as functions of $\widetilde{\Delta}/\Gamma$ and $\widetilde{\delta}/\Gamma$
for $C=16$, $N_{1}=N_{2}=4$, and $\Omega/\Gamma=1$. The effective
decay rates are $(\Gamma_{1},\Gamma_{2})/\Omega=(1/4,1/4)$ with $(\widetilde{\phi}_{a},\widetilde{\phi}_{b})/\pi\simeq(9/16,9/16)$
{[}panels (a)--(c){]}, $(\Gamma_{1},\Gamma_{2})/\Omega=(1/2,1/8)$
with $(\widetilde{\phi}_{a},\widetilde{\phi}_{b})/\pi\simeq(3/5,16/17)$
{[}panels (d)--(f){]}, and $(\Gamma_{1},\Gamma_{2})/\Omega=(1/8,1/2)$
with $(\widetilde{\phi}_{a},\widetilde{\phi}_{b})/\pi\simeq(16/17,3/5)$
{[}panels (g)--(i){]}. The white dashed lines indicate $\widetilde{\Delta}=0$
and $\widetilde{\Delta}=\widetilde{\delta}$, respectively.}
\end{figure}

In the strong coupling regime, we examine the cases with $(\Gamma_{1},\Gamma_{2})/\Omega=(1/4,1/4)$
{[}$(\widetilde{\phi}_{a},\widetilde{\phi}_{b})/\pi\simeq(9/16,9/16)$,
Figs. \ref{Fig.4 Strong}(a)--\ref{Fig.4 Strong}(c){]}, $(\Gamma_{1},\Gamma_{2})/\Omega=(1/2,1/8)$
{[}$(\widetilde{\phi}_{a},\widetilde{\phi}_{b})/\pi\simeq(3/5,16/17)$,
Figs. \ref{Fig.4 Strong}(d)--\ref{Fig.4 Strong}(f){]}, and $(\Gamma_{1},\Gamma_{2})/\Omega=(1/8,1/2)$
{[}$(\widetilde{\phi}_{a},\widetilde{\phi}_{b})/\pi\simeq(16/17,3/5)$,
Figs. \ref{Fig.4 Strong}(g)--\ref{Fig.4 Strong}(i){]}, respectively.
Here, for $\Gamma_{1}<\Omega,\Gamma_{2}<\Omega$, and $C\gg1$, anti-crossings
centered at $\widetilde{\Delta}=\widetilde{\delta}=0$ emerge in the
scattering probabilities $T_{1\rightarrow2}$, $R_{1\rightarrow1}$,
and $T_{1\rightarrow3(4)}$. The branch widths are set by the effective
decay rates $\Gamma_{1}$ and $\Gamma_{2}$, while their splitting
is a direct signature of the interatomic-coupling-induced mode hybridization.
As a result, the transmission $T_{1\rightarrow2}$ exhibits a distinct
double\nobreakdash-dip structure {[}Figs. \ref{Fig.4 Strong}(a),
\ref{Fig.4 Strong}(d), \ref{Fig.4 Strong}(g){]}, and the reflection
$R_{1\rightarrow1}$ exhibits a double\nobreakdash-peak structure
{[}Figs. \ref{Fig.4 Strong}(b), \ref{Fig.4 Strong}(e), \ref{Fig.4 Strong}(h){]},
as functions of $\widetilde{\Delta}$, with the separation between
dips (or peaks) growing with $|\widetilde{\delta}|$. At the center
$\widetilde{\Delta}=\widetilde{\delta}=0$, the reflection drops to
a minimum $R_{1\rightarrow1}=(1+C)^{-2}$. Conversely, when $|\widetilde{\delta}|$
becomes sufficiently large, the two atoms effectively decouple, and
the reflection rises sharply near $\widetilde{\Delta}=0$.

The transmission probability to waveguide\,B, $T_{1\rightarrow3(4)}$,
shown in Figs. \ref{Fig.4 Strong}(c), \ref{Fig.4 Strong}(f) and
\ref{Fig.4 Strong}(i), exhibits two peaks with a maximum value of
0.25, each located on one of the branches. Specifically, their positions
in the ($\widetilde{\Delta},\widetilde{\delta}$) plane are given
by \{$\widetilde{\Delta}=\Gamma_{1}\sqrt{C-1},$ $\widetilde{\delta}=(\Gamma_{1}-\Gamma_{2})\sqrt{C-1}\}$
and \{$\widetilde{\Delta}=-\Gamma_{1}\sqrt{C-1},$ $\widetilde{\delta}=(\Gamma_{2}-\Gamma_{1})\sqrt{C-1}\}$.
Notably, only when the decay rates are symmetric ($\Gamma_{1}=\Gamma_{2}$,
i.e., $\mathcal{R}=1$), do the resonant peaks lie along $\widetilde{\delta}=0$,
at $\widetilde{\Delta}=\pm\Gamma_{1}\sqrt{C-1}$. When the asymmetry
is strong---e.g., for $\Gamma_{1}/\Gamma_{2}=4$ ($\mathcal{R}>1$)
and $\Gamma_{2}/\Gamma_{1}=4$ ($\mathcal{R}<1$)---the maxima of
$T_{1\rightarrow3(4)}$ shift toward the tails closer to the line
$\widetilde{\Delta}=\widetilde{\delta}$ and to the line $\widetilde{\Delta}=0$,
respectively, rather than remaining along $\widetilde{\delta}=0$.
Physically, the process can be understood as follows: The photon enters
waveguide\,A, couples to atom\,1, transfers coherently to atom\,2,
and radiates into waveguide\,B. Crucially, this involves all possible
scattering paths---including multiple back\nobreakdash-and\nobreakdash-forth
exchanges between the two atoms. This coherent superposition is equivalent
to exciting the two hybrid eigenmodes of the system, which then radiate
into the output. Due to the asymmetric decay rates of the two atoms
($\Gamma_{1}\neq\Gamma_{2}$), their phase responses to frequency
differ, leading to a relative phase shift between the fields radiated
by the two hybrid modes into the output waveguide, which can result
in destructive interference. To maximize transmission, the detuning
$\widetilde{\delta}$ between the two atoms and the probe detuning
$\widetilde{\Delta}$ must be adjusted to compensate for the phase
delay introduced by the decay\nobreakdash-rate asymmetry, so that
the fields radiated by the two hybrid modes add in phase at the output,
achieving constructive interference. Mathematically, the optimum transmission
condition can be obtained by solving $\partial T_{1\rightarrow3(4)}/\partial\widetilde{\Delta}$
and $\partial T_{1\rightarrow3(4)}/\partial\widetilde{\delta}$. This
yields the relations $\widetilde{\delta}=\left(1-\mathcal{R}^{-1}\right)\widetilde{\Delta}$,
$\widetilde{\Delta}^{2}=\Gamma_{1}^{2}(C-1)$, which together determine
the operating point of maximum transmission into waveguide\,B.

\section{The Non-Markovian regime\label{sec:4}}

We now consider the non-Markovian regime,
where the propagation delay $\tau^{a(b)}$ between coupling points
becomes comparable to the characteristic atomic relaxation time $\sim1/\left[N_{1(2)}^{2}\Gamma\right]$ \citep{Du2021,Chen2022,Zhou2023,li2024single,zhu2025single,Lim2023,Sathyamoorthy2014}.
In this regime, the phase corrections $|\Delta|\tau^{a}$ and $|\Delta-\delta|\tau^{b}$---which
add to the static phases $\widetilde{\phi}_{a}$ and $\widetilde{\phi}_{b}$---must
be carefully included because $\{|\Delta|,|\Delta-\delta|\}$ are
typically of order $\Gamma$. These corrections can significantly
modify the Lamb shifts $\Delta_{\text{ls},1(2)}$, thereby shifting
the transmission peak (or the avoided\nobreakdash-crossing centers)
for\textcolor{red}{{} }$C<1$ (or $C>1$) in a periodic fashion. Moreover,
the effective decay rates $\Gamma_{1}$ and $\Gamma_{2}$ are also
strongly altered, causing the cooperativity parameter $C\sim1/(\Gamma_{1}\Gamma_{2})$
to vary between values below and above unity. As a result, non\nobreakdash-Markovian
retardation can induce transitions between the weak\nobreakdash- and
strong\nobreakdash-coupling regimes. This intriguing behavior can
be illustrated by a simple example with $\widetilde{\phi}_{a}=\widetilde{\phi}_{b}$
and $\tau^{a}=\tau^{b}\eqqcolon\tau$.

\subsection{Non-Markovianity induced transition from weak to strong coupling
regime: $C<1\rightarrow C>1$}

\begin{figure}
\includegraphics[width=1\columnwidth]{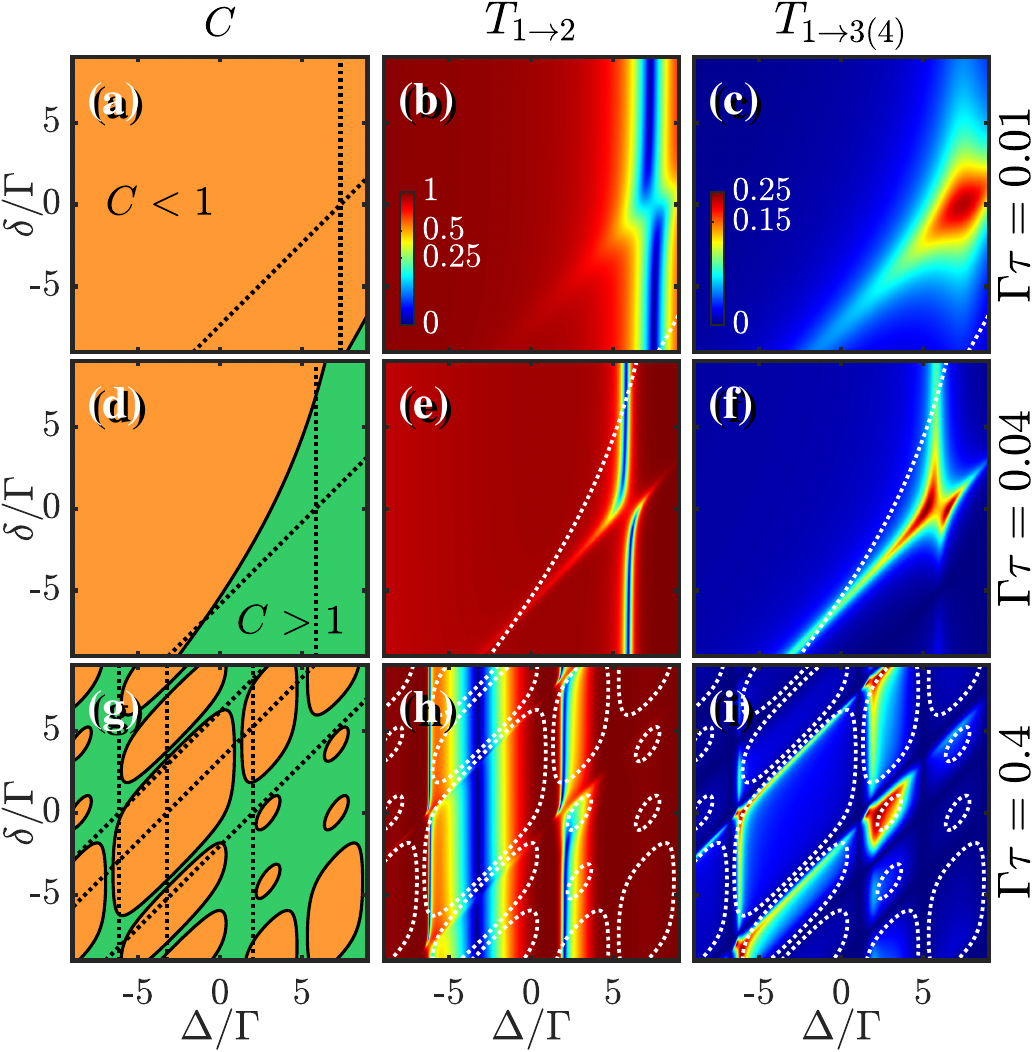}

\caption{\label{Fig.5Weak_Non}\textbf{Non-Markovianity induced transition
from weak to strong coupling regime.} Cooperativity parameter $C$
and transmission probabilities $T_{1\rightarrow2}$ and $T_{1\rightarrow4}$
versus $\Delta/\Gamma$ and $\delta/\Gamma$ in the weak coupling
regime. The propagation delay between neighbouring coupling points
are $\Gamma\tau=0.01$ {[}panels (a)--(c){]}, $\Gamma\tau=0.04$
{[}panels (d)--(f){]}, and $\Gamma\tau=0.4$ {[}panels (g)--(i){]}.
Other parameters are the same as those in Fig. \ref{Fig.2Weak}(a).
In panels (a), (d), and (g), the black dotted lines indicate the $\Delta=\Delta_{\text{ls},1}$
and $\Delta-\delta=\Delta_{\text{ls},2}$. In the remaining panels,
the white dotted lines mark the boundary between the $C>1$ and $C<1$
regimes.}
\end{figure}

We first examine the weak-coupling regime shown in Figs.{
\ref{Fig.2Weak}(a)--\ref{Fig.2Weak}(c)} under the Markovian approximation. Fig.\,\ref{Fig.5Weak_Non}
presents the cooperativity parameter $C$ and the transmission probabilities
$T_{1\rightarrow2}$ and $T_{1\rightarrow3(4)}$ as functions of the
probe detuning $\Delta$ and the atomic detuning $\delta$, for
three values of the delay time $\Gamma\tau=\{0.01,0.04,0.4\}$, with
$N_{1}=N_{2}=4$, $\Omega/\Gamma=1$ and $(\widetilde{\phi}_{a},\widetilde{\phi}_{b})/\pi\simeq(9/25,9/25)$.
In the phase diagrams for $C$, the strong\nobreakdash-coupling regime
($C>1$) is shaded green and the weak\nobreakdash-coupling regime
($C<1$) is shaded orange. The black dotted lines indicate the resonance
conditions $\Delta=\Delta_{\text{ls},1}$ and $\Delta-\delta=\Delta_{\text{ls},2}$.
This same color and line scheme is used consistently in the following
figures.

For $\Gamma\tau=0.01$ {[}Figs. \ref{Fig.5Weak_Non}(a)--\ref{Fig.5Weak_Non}(c){]},
the system predominantly remains in the weak\nobreakdash-coupling
regime ($C<1$), with only a small parameter region turning into the
strong\nobreakdash-coupling regime ($C>1$) {[}Fig. \ref{Fig.5Weak_Non}(a){]}.
Consequently, the transmission spectra in Figs. \ref{Fig.5Weak_Non}(b)
and \ref{Fig.5Weak_Non}(c) remain close to the Markovian limit: $T_{1\rightarrow2}$
exhibits two slightly separated valleys near the intersection of the
resonances $\Delta=\Delta_{\text{ls},1}$ and $\Delta-\delta=\Delta_{\text{ls},2}$,
while $T_{1\rightarrow4}$ retains a single, unsplit peak. In this
weak non\nobreakdash-Markovian regime, the phase corrections $\Delta\tau$
and $(\Delta-\delta)\tau$ typically modify the effective decay rates
$\Gamma_{1}$ and $\Gamma_{2}$ only slightly. However, when $\Delta$
and $\delta$ have opposite signs, the phase correction $|\Delta-\delta|\tau$
(approximately $|2\Delta|\tau$ near the corner) may become predominant,
causing $\Gamma_{1(2)}$ to decrease sharply. This reduction can push
the cooperativity $C$ across the threshold from $C<1$ to $C>1$,
as seen in the lower\nobreakdash-right region.

Increasing the delay to $\Gamma\tau=0.04$ extends the strong\nobreakdash-coupling
regime ($C>1$) across approximately half of the $(\Delta,\delta)$
parameter space {[}Fig. \ref{Fig.5Weak_Non}(d){]}. The stronger phase
corrections $\Delta\tau$ and $(\Delta-\delta)\tau$ now modify $\Gamma_{1}$
and $\Gamma_{2}$ more prominently, pushing more of parameter space
into the strong-coupling regime. In this case, the resonant intersection
now falls within the $C>1$ region, resulting in clear anti-crossings
in $T_{1\rightarrow2}$ {[}Fig. \ref{Fig.5Weak_Non}(e){]} and $T_{1\rightarrow4}$
{[}Fig. \ref{Fig.5Weak_Non}(f){]}. Furthermore, the asymmetry in
the phase corrections {[}$\Delta\tau\neq(\Delta-\delta)\tau${]} can
effectively invert the relative magnitudes of the decay rates, transitioning
from $\Gamma_{1}>\Gamma_{2}$ to $\Gamma_{2}<\Gamma_{1}$, thereby
slightly distorting the transmission spectra.

With a time delay of $\Gamma\tau=0.4$, the phase diagram exhibits
a multiple\nobreakdash-resonance landscape: discrete islands of weak
coupling ($C<1$) surrounded by an extended strong\nobreakdash-coupling
($C>1$) region {[}Fig. \ref{Fig.5Weak_Non}(g){]}. Because the phase
corrections $\Delta\tau$ and $(\Delta-\delta)\tau$ change rapidly
with $\Delta$ or $\delta$ in this highly non\nobreakdash-Markovian
regime (large $\tau$), the Lamb shifts $\Delta_{\text{ls},1(2)}$
and decay rates $\Gamma_{1(2)}$ exhibit pronounced periodic dependence.
This results in the resonance conditions $\Delta=\Delta_{\text{ls},1}$
and $\Delta=\delta+\Delta_{\text{ls},2}$ being satisfied at multiple
points, with the corresponding $C$ transiting between values below
and above unity. As shown in Figs. \ref{Fig.5Weak_Non}(h) and \ref{Fig.5Weak_Non}(i),
both $T_{1\rightarrow2}$ and $T_{1\rightarrow4}$ exhibit multiple anti-crossings
at the resonant intersections when $C>1$, whereas they retain a single\nobreakdash-peak
profile when $C<1$.\textcolor{red}{{} }When the system parameters lie
along $\Delta/\Gamma\sim-3$ and approach the resonance conditions,
the large effective decay rates $\Gamma_{1(2)}$ suppress the cooperativity
($C\rightarrow0$), so that no resolvable structures emerge in the
spectrum.

\subsection{Non-Markovianity induced transition from strong to weak coupling
regime: $C>1\rightarrow C<1$}

Conversely, non\nobreakdash-Markovian retardation can also induce
transitions from the strong\nobreakdash- to the weak\nobreakdash-coupling
regime. To illustrate this, we examine the strong\nobreakdash-coupling
regime previously shown in Figs. {\ref{Fig.4 Strong}(a)}--{\ref{Fig.4 Strong}(c)}, with
$N_{1}=N_{2}=4$, $\Omega/\Gamma=1$ and $(\widetilde{\phi}_{a},\widetilde{\phi}_{b})/\pi\simeq(9/16,9/16)$.
For $\Gamma\tau=0.04$, consistent with the preceding analysis, incorporating
non-Markovian delay effects modifies $C$, resulting in localized
regions where $C>1$ switches to $C<1$ {[}Fig. \ref{Fig.6Strong_Non}(a){]}.
Here, the anti-crossing emerges closer to the boundary separating
the weak\nobreakdash- and strong\nobreakdash-coupling regions. Compared
to the corresponding Markovian case [Figs. {\ref{Fig.4 Strong}(a)}, {\ref{Fig.4 Strong}(c)], the gap of the anti-crossing
is visibly reduced, as seen in Figs. \ref{Fig.6Strong_Non}(b) and
\ref{Fig.6Strong_Non}(c).

For $\Gamma\tau=0.24$, the weak coupling region expands and multiple
weak-coupling ``islands'' emerge {[}Fig. \ref{Fig.6Strong_Non}(d){]}.
As seen in Figs. \ref{Fig.6Strong_Non}(e) and \ref{Fig.6Strong_Non}(f),
the resonant intersection now lies within a weak\nobreakdash-coupling
island. Consequently, the avoided crossing (that would appear in the
Markovian regime) closes, owing to the non-Markovianity-induced
transition from $C>1$ to $C<1$. Moreover, the pronounced peak in
Fig. \ref{Fig.6Strong_Non}(f) indicates that $T_{1\rightarrow4}$
approaches the maximum value of 0.25, in contrast to the typical transmission
$T_{1\rightarrow4}<0.25$ observed in the weak\nobreakdash-coupling
Markovian regime [see Fig. \ref{Fig.2Weak}(c)].

In the deep non\nobreakdash-Markovian regime with $\Gamma\tau=0.4$
{[}Figs. \ref{Fig.6Strong_Non}(g)--\ref{Fig.6Strong_Non}(i){]},
the detuning-dependent phases vary rapidly so that the cooperativity
parameter $C$ oscillates repeatedly between $C>1$ and $C<1$ across
the finite parameter range. This rapid oscillation effectively partitions
the parameter space into an alternating mosaic of strong- and weak\nobreakdash-coupling
regions. The transmission spectra $T_{1\rightarrow2}$ and $T_{1\rightarrow4}$
directly reflect this partitioning: in regions where $C>1$, the spectrum
exhibits clear anti-crossings, while for resonant intersection points
lying within a $C<1$ region, the transmission features coalesce into
a single, broadened peak. In the limit $C\rightarrow0$, these features
become vanishingly small.
\begin{figure}
\includegraphics[width=1\columnwidth]{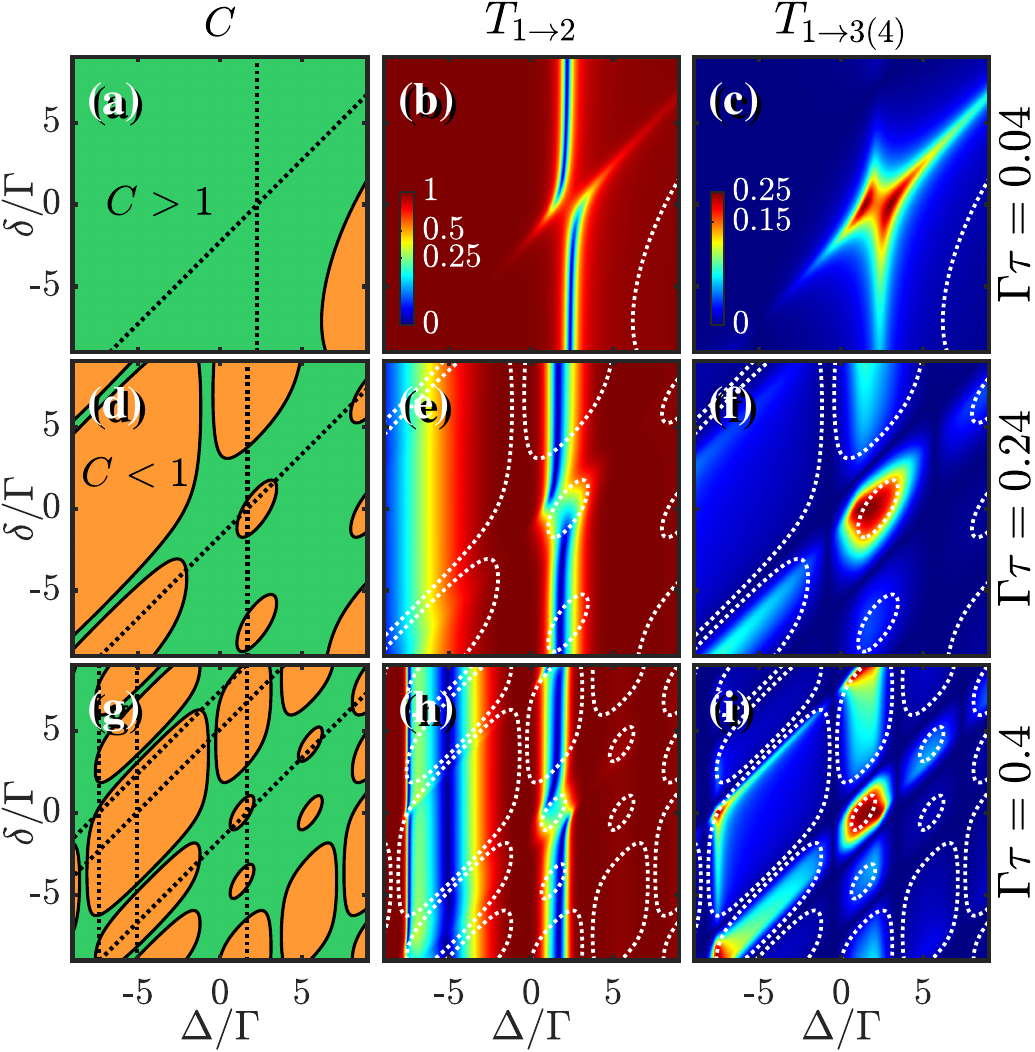}

\caption{\label{Fig.6Strong_Non}\textbf{Non-Markovianity induced transition
from strong to weak coupling regime.} Cooperativity parameter $C$
and transmission probabilities $T_{1\rightarrow2}$ and $T_{1\rightarrow3(4)}$
versus $\Delta/\Gamma$ and $\delta/\Gamma$ in the strong coupling
regime, with $\Gamma\tau=0.04$ {[}panels (a)--(c){]}, $\Gamma\tau=0.24$
{[}panels (d)--(f){]}, and $\Gamma\tau=0.4$ {[}panels (g)--(i){]},
respectively. Other parameters are the same as those in Fig. \textcolor{red}{\ref{Fig.4 Strong}}(a).
In panels (a), (d), and (g), the black dotted lines indicate the $\Delta=\Delta_{\text{ls},1}$
and $\Delta-\delta=\Delta_{\text{ls},2}$. In the remaining panels,
the white dotted lines mark the boundary between the $C>1$ and $C<1$
regimes.}
\end{figure}

\section{DISCUSSION AND EXPERIMENTAL FEASIBILITY\label{sec:5}}

We further investigate the chiral coupling mechanism\citep{li2024single,Zhou2023,BelloXXXX,Chen2022,zheng2024chiral,sun2025phase},
where the coupling strengths between atoms and left- and right-propagating
optical fields in the waveguide are asymmetric ($\Gamma_{1L}\neq\Gamma_{1R}$,
$\Gamma_{2L}\neq\Gamma_{2R}$). This chiral interaction breaks the
spatial symmetry of the system, leading to directional transfer of
photons. To clearly reveal its core physical effect, we focus on the
ideal chiral coupling condition ($\Gamma_{1L}=\Gamma_{2L}=0$, $\Gamma_{1R}=\Gamma_{1}$,
$\Gamma_{2R}=\Gamma_{2}$). Under this condition, atoms only couple
to right-propagating photons, resulting in no photon output at port
1 (reflection) or port 3 (transmission), i.e., $r_{1\rightarrow1}=t_{1\rightarrow3}=0$,
the transmission amplitudes simplify to
\begin{align}
t_{1\rightarrow2} & =1+\frac{2i\Gamma_{1}[(\widetilde{\Delta}-\widetilde{\delta})+i\Gamma_{2}]}{\Omega^{2}-(\widetilde{\Delta}+i\Gamma_{1})[(\widetilde{\Delta}-\widetilde{\delta})+i\Gamma_{2}]},\nonumber \\
t_{1\rightarrow4} & =\frac{2i\sqrt{\Gamma_{1}\Gamma_{2}}\Omega}{\Omega^{2}-(\widetilde{\Delta}+i\Gamma_{1})[(\widetilde{\Delta}-\widetilde{\delta})+i\Gamma_{2}]},\label{eq:chiral_t12}
\end{align}
where $\widetilde{\Delta}=\Delta-\Delta_{\text{ls},1}$ and $\widetilde{\delta}=\delta-\Delta_{\text{ls},1}+\Delta_{\text{ls},2}$.
According to energy conservation, the system satisfies $T_{1\rightarrow2}+T_{1\rightarrow4}=1$.

Figure \ref{Fig7Chiral} shows the variation of $T_{1\rightarrow4}$
with $\Delta$ and $\delta$ in the Markovian and non-Markovian regimes.
Compared to the non-chiral case shown in Fig. \ref{Fig.4 Strong},
the most striking difference is that under specific detuning conditions
\{$\widetilde{\Delta}=\Gamma_{1}\sqrt{C-1},$ $\widetilde{\delta}=(\Gamma_{1}-\Gamma_{2})\sqrt{C-1}\}$
and \{$\widetilde{\Delta}=-\Gamma_{1}\sqrt{C-1},$ $\widetilde{\delta}=(\Gamma_{2}-\Gamma_{1})\sqrt{C-1}\}$,
the system achieves perfect transmission of photons to port 4 ($T_{1\rightarrow4}=1$).

This phenomenon can be understood in terms of quantum interference
between different transmission paths. The transmission amplitude from
port 1 to port 2 can be regarded as the coherent superposition of
two paths: one is the direct background transmission path (corresponding
to the constant \textquotedbl 1\textquotedbl{} in the expression
for $t_{1\rightarrow2}$), and the other is the resonant path where
the photon is coherently transferred via the hybridized modes of the
giant molecule. In the non-chiral case ($\Gamma_{1L}=\Gamma_{1R}$,
$\Gamma_{2L}=\Gamma_{2R}$), the symmetry of atomic decay causes partial
destructive interference between the field radiated from the resonant
path and the direct path, resulting in a transmission amplitude of
$t_{1\rightarrow2}=1-\Gamma_{1R}/\Gamma_{1}=1/2$. However, under
ideal chiral coupling conditions ($\Gamma_{1L}=\Gamma_{2L}=0$, $\Gamma_{1R}=\Gamma_{1}$,
$\Gamma_{2R}=\Gamma_{2}$) the unidirectional decay significantly
enhances the coupling efficiency of the resonant path, increasing
its amplitude to 1, thereby leading to complete destructive interference
with the direct path and fully suppressing the output at port 2 ($t_{1\rightarrow2}=1-\Gamma_{1R}/\Gamma_{1}=0$).

In the non-Markovian regime {[}Figs. \ref{Fig7Chiral}(d)--\ref{Fig7Chiral}(f){]},
retardation effects cause the system behavior to transition between
the strong- and weak-coupling regimes. Compared with the non-chiral
case, when the photon is resonant with both atoms (corresponding to
the intersection of the black dotted-dashed lines in the figure):
if the system is in the strong-coupling regime, the spectrum of $T_{1\rightarrow4}$
exhibits splitting with a peak value of 1; if the system is in the
weak-coupling regime, the spectral splitting disappears and the maximum
value is less than 1.

The giant-atom model studied here is directly relevant to several
recent superconducting-circuit experiments. Multi-point giant atoms
with three and six coupling points have been realized, demonstrating
phenomena such as decoherence-free interactions and electromagnetically
induced transparency \citep{Vadiraj2021,Kannan2020}. Moreover, circuit-QED
platforms routinely employ frequency-tunable transmons and arrays
of detuned qubits with engineered couplings, enabling the creation
of dressed-state spectra and photonic bound states \citep{shen2007strongly,wang2024controlling,McKay2016,Stehlik2021}.

\begin{figure}
\includegraphics[width=1\columnwidth]{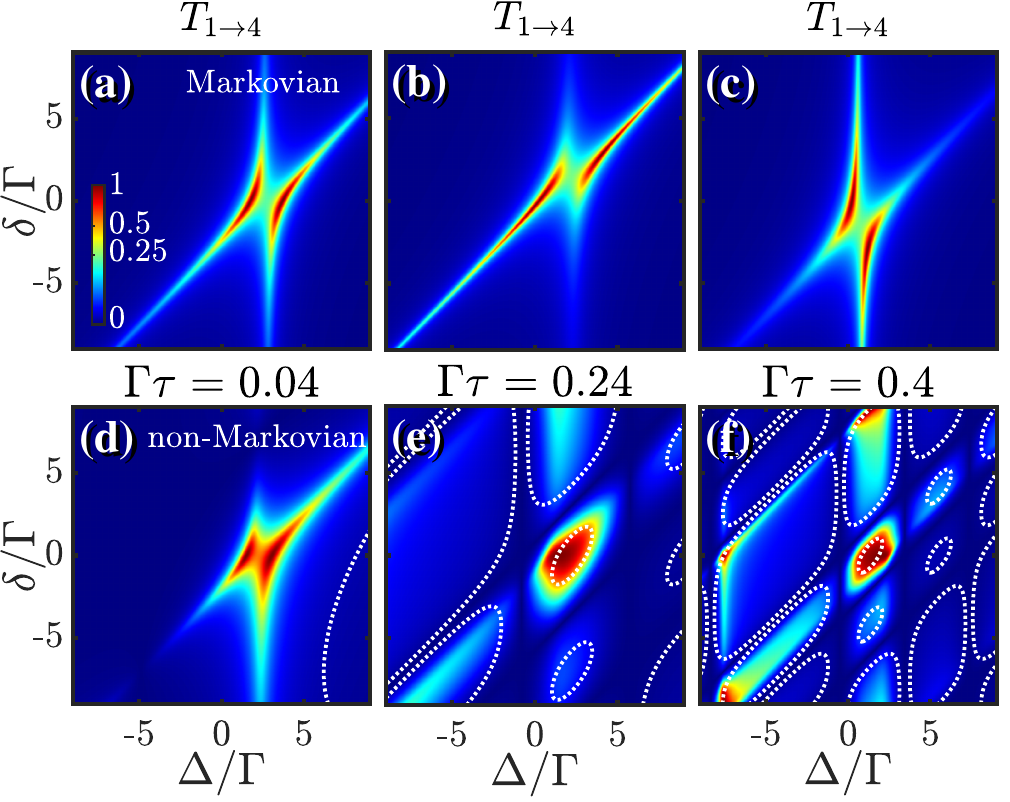}\caption{\label{Fig7Chiral}\textbf{Chirality-induced directional scattering.}
Transmission probabilities $T_{1\rightarrow4}$ as functions of $\Delta/\Gamma$
and $\delta/\Gamma$ under ideal chiral coupling in the strong coupling
regime. (a)--(c) show $T_{1\rightarrow4}$ in the Markovian regime,
corresponding to three different effective decay rates (as in Fig.
\ref{Fig.4 Strong}). (d)--(f) display $T_{1\rightarrow4}$ in the
Non-Markovian regime for $\Gamma\tau=0.04$, $\Gamma\tau=0.24$, and
$\Gamma\tau=0.4$, respectively. Other parameters are the same as
those in Fig. \textcolor{red}{\ref{Fig.4 Strong}}(a). In the panel
(d)--(f), the white dotted lines mark the boundary between the $C>1$
and $C<1$ regimes.}
\end{figure}

A closely related key experimental demonstration was reported in \citep{Kannan2023,Almanakly2025},
where an artificial molecule comprising two superconducting qubits
was strongly coupled to a bidirectional waveguide.
In this setup, each qubit couples to the waveguide with a strength
of $2\pi\times3$ MHz, while a tunable coupler provides a controllable
exchange interaction between the qubits, tunable from $2\pi\times1$
MHz to $2\pi\times10$ MHz. The frequency detuning between the qubits,
as well as between the probe and the atomic resonances, can be varied
over a range of a few to several tens of MHz. Furthermore, qubit-waveguide
coupling strengths can reach $2\pi\times17$ MHz \citep{Almanakly2025},
and can be further increased with optimized designs such as superconducting
flux qubits \citep{Kannan2023}. The physical separation between the
two atoms can, in principle, be arbitrarily large; separations up
to 10\,cm have already been demonstrated \citep{Almanakly2025}. In
addition, chiral quantum optics has also been achieved in microwave
settings, typically using ferrite circulators to break reciprocity
in transmission lines \citep{Sathyamoorthy2014,Sliwa2015,Chapman2017,Mueller2018}.
Since the giant artificial atoms realized in \citep{Kannan2020,Kannan2023,Almanakly2025,Vadiraj2021}
are coupled to microwave transmission lines, deterministic photon
routing via chirally coupled giant atoms could be implemented by incorporating
such circulators to introduce the required directionality. Taken together,
these experimental advances confirm that the giant molecule
configurations with a tunable atomic detuning analyzed in this work are well within reach of current
superconducting-circuit technology. In parallel, experiments coupling
superconducting qubits to SAW modes have explored the deeply non-Markovian
regime by using coupling points separated by hundreds of wavelengths.
These SAW-based systems provide an alternative platform for observing the effects predicted here.

\section{CONCLUSION\label{sec:6}}
In conclusion, we have investigated single-photon scattering in a waveguide quantum electrodynamics system consisting of a giant molecule formed by two frequency-detuned giant atoms. Each atom is coupled to one of two parallel waveguides via multiple spatially separated connection points, and the atoms are coherently coupled to each other through a tunable exchange interaction. By introducing the cooperativity parameter $C$, we classify the scattering behavior into three distinct regimes: weak ($C<1$), critical ($C=1$), and strong ($C>1$) coupling.

In the Markovian regime, where propagation delays between coupling points are negligible, we demonstrate that the scattering probabilities can be precisely controlled by tuning the atomic detuning and the probe detuning. The competition between coherent atom--atom coupling and radiative decay into the waveguides gives rise to rich spectral features, including Lorentzian and anti-Lorentzian lineshapes, electromagnetically induced transparency windows, and avoided crossings. The optimal scattering probabilities between the waveguides are jointly determined by the atomic detuning, the effective decay rates, and their asymmetry ratio, emerging from path interference.

When non-Markovian delays become significant, we show that even moderate time delays can expand or fragment the weak- or strong-coupling regions in parameter space, actively inducing transitions between these regimes. In the deep non-Markovian regime, this behavior manifests as multiple peaks and avoided crossings in the scattering spectra. We further show that the interference-based photon routing can be rendered deterministic by incorporating chiral atom-waveguide couplings. The interplay between detuning, retardation, and self-interference thus provides a versatile toolbox for photon-level network control and opens new avenues for exploring non-Markovian quantum optics in tailored photonic environments.

\begin{acknowledgments}
H.W. acknowledges support from the National Natural Science Foundation
of China under Grant No. 12174058. Y.L. was supported by the National
Natural Science Foundation of China under Grant No. 12274107 and the
Hainan Provincial Natural Science Foundation of China (Grant No. 125RC631).
\end{acknowledgments}

\bibliographystyle{apsrev4-2}
\bibliography{ref251209}

\end{document}